\documentclass[aps,prl,twocolumn,superscriptaddress]{revtex4}
\usepackage{ae}
\usepackage[T1]{fontenc}
\usepackage[ansinew]{inputenc}
\usepackage{amsmath}
\usepackage{amssymb}
\usepackage{graphicx}
\usepackage{color}

\usepackage[colorlinks]{hyperref}
\usepackage{epstopdf}

\def\aa#1{\textcolor{blue}{#1}}
\def\sig{{\mbox{\boldmath{$\sigma$}}}}

\def\sig{{\mbox{\boldmath{$\sigma$}}}}
\usepackage{soul,xcolor}
\setstcolor{red}                   % Change color of strike-through line created by "\st" here.

\def\sig{{\mbox{\boldmath{$\sigma$}}}}

\begin{document}

\date{\today}

\title{ Photovoltaic effect generated by spin-orbit interactions}

\author{O. Entin-Wohlman}
\email{entin@tau.ac.il}
\affiliation{Raymond and Beverly Sackler School of Physics and Astronomy, Tel Aviv University, Tel Aviv 69978, Israel}
%\affiliation{Physics Department, Ben Gurion University, Beer Sheva 84105, Israel}

\author{R. I. Shekhter}
\affiliation{Department of Physics, University of Gothenburg, SE-412
96 G{\" o}teborg, Sweden}

\author{M. Jonson}
\affiliation{Department of Physics, University of Gothenburg, SE-412
96 G{\" o}teborg, Sweden}

\author{A. Aharony}
\affiliation{Raymond and Beverly Sackler School of Physics and Astronomy, Tel Aviv University, Tel Aviv 69978, Israel}

\begin{abstract}
An AC electric field applied to a junction comprising two spin-orbit coupled weak links connecting a quantum dot to two electronic terminals is proposed to  induce a DC current and to generate  a voltage drop over the junction if it is  a part of an open circuit. This photovoltaic effect requires a junction in which mirror reflection-symmetry is broken.  Its origin lies in the different fashion inelastic processes modify the reflection of electrons from the junction back into the two terminals, which leads to uncompensated DC transport. The  effect can be detected by measuring the  voltage drop  that is built up due to that DC  current.
This voltage is an  even function  of the frequency of the AC electric field. % and  is {\em not} related to quantum pumping.

\end{abstract}

\maketitle

%%%%%%%%%%%%%%%%%%%%%%%%%%%%%%%%%%%%%%%%%%%%%%%%%%%%%%
%%%%%%%%%%%%%%%%%%%%%%%%%%%%%%%%%%%%%%%%%%%%%%%%%%%%%%
%\section{Introduction}
%\label{intro}

\noindent{\bf {\it 1. Introduction.}} Electric weak links made of materials with strong spin-orbit interactions open a promising
way to achieve spin-dependent transport of electrons.
In the particular case of the Rashba spin-orbit coupling  \cite{Rashba},  the interaction can be tuned electrostatically \cite{Nitta,Sato,Beukman}
or mechanically \cite{RS2013, MJ2018}.   This coupling  obeys time-reversal symmetry which prevents  spin splitting of electron transport in  two-terminal junctions \cite{bardarson},   in most cases eliminating the possibility   to manipulate  electronic conduction through  Rashba weak links.
Spin-orbit interactions do, however, have an effect on spin-polarized electrons in magnetic materials \cite{datta,sarkar,
RS2014},  and on electrons subjected to
external magnetic fields \cite{lyanda,janine,us,Saarikoski,Shmakov,Nagasawa}.
Here we propose that imposing a {\it time dependence} on the effective magnetic fields induced by the spin-orbit coupling  offers another means to destroy time-reversal symmetry  of two-terminal junctions. In particular we predict that  time-dependent Rashba interactions      generate a DC electric current through unbiased junctions.

Coherent electronic transport in response to   periodic modulations
of the shape of   quantum dots or of other potential parameters of mesoscopic junctions
has been attracting considerable  interest
 \cite{Altshuler,Avron} following the seminal paper by Thouless  \cite{Thouless},  who showed that a slow periodic variation of the potential landscape may yield quantized and non-dissipative particle transport in unbiased junctions--a phenomenon termed ``adiabatic quantum pumping".
Adiabatic pumping of spin currents  resulting from periodic modulations of the shape of a spin-orbit coupled junction  has been discussed as well \cite{Sharma},
also as a result of temporal modulations of the Rashba interaction \cite{Governale,Avishai,Brosco,MJ2019}.  However, the possibility to induce a DC {\em particle current} by such modulations in the absence of a bias voltage was not considered.

%%%%%%%%%%%%%%%%%%%%%%%%%%%%%%%%%%%%%%%%%%%%%%%%%%%%%%%%%%%%
\begin{figure}[htp]
\includegraphics[width=8.5cm]{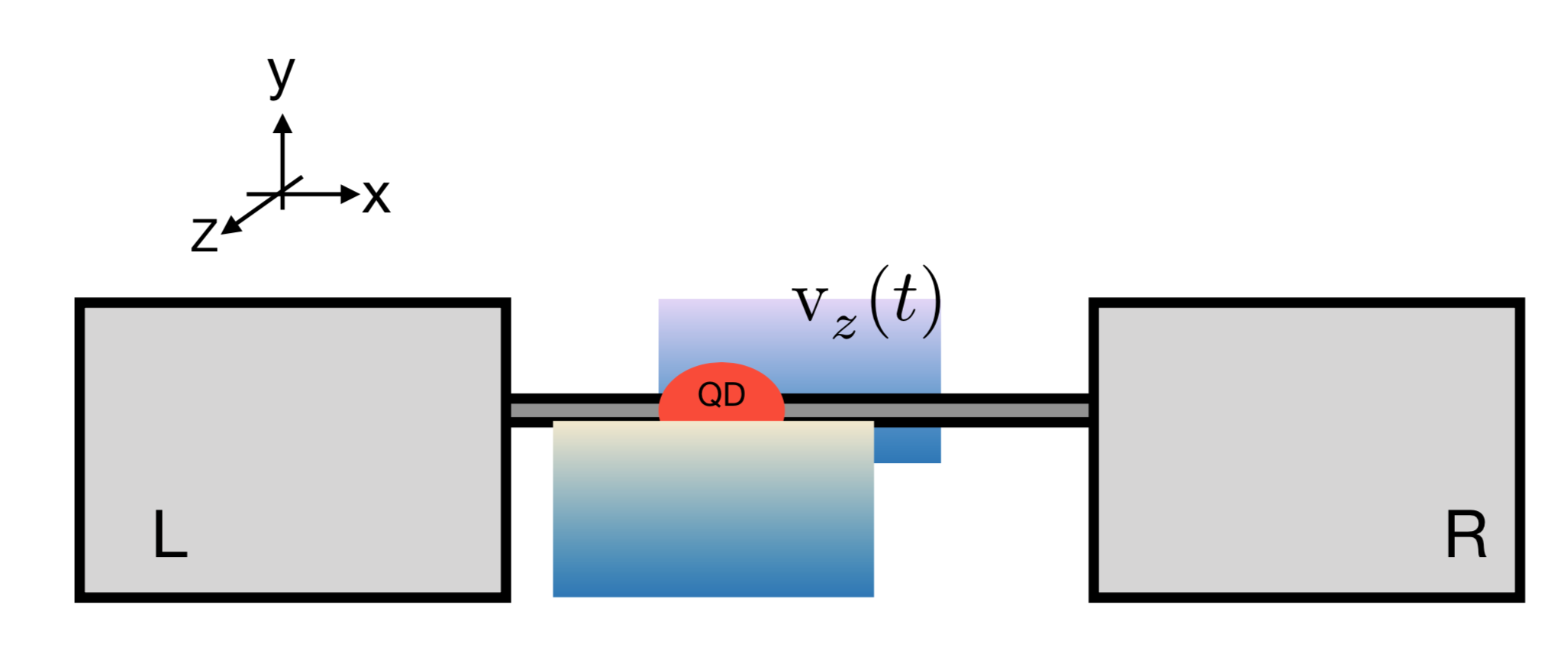}
\caption{(Color online.)  Illustration of the model system. A quantum dot, represented by a localized energy level, is attached by two weak links  lying in the $x-y$ plane  to two reservoirs, denoted $L$ and $R$. An AC electric field along %the direction of
$\hat{\bf z}$, whose amplitude oscillates with frequency $\Omega$, induces a Rashba spin-orbit interaction in the links. }%which leads to time-dependent spin-flips as electrons pass through the junction.}%The bent in the wire is introduced for generality--the bent angle does not affect the particle current, but  does modify spin currents.
\label{sys}
\end{figure}
%%%%%%%%%%%%%%%%%%%%%%%%%%%%%%%%%%%%%%%%%%%%%%%%%%%%%%%%%%%%

DC charge transport driven by time-dependent spin-orbit coupling is an alternative to the pumping of charge  caused by  tuning periodically  the potential landscape of mesoscopic structures. It occurs in
inhomogeneous junctions in which mirror reflection-symmetry is violated.  In an unbiased  junction   no net current flows when the spin-orbit interaction is static, even in an asymmetric device: transport of electrons incident from the two opposite reservoirs is   fully equilibrated. In fact, a static spin-orbit coupling,  which results in a unitary evolution of the spinor wave-function,  does not modify the DC transport. However,  unitarity
 is destroyed  by the time dependence   that entails additional reflection processes due to inelastic tunneling.  These in general differ
for  the two opposite directions in which electrons can be reflected from the junction, leading to uncompensated electronic transport. To elaborate on this general statement  we refer to the device illustrated in Fig. \ref{sys}: a quantum dot represented by a single level of energy $\epsilon$ is
connected by  spin-orbit coupled weak links to left and right reservoirs.  Due to the Aharonov-Casher effect \cite{AC}, the tunneling matrix elements attain   unitary-matrix (in spin space) phase factors \cite{Oreg}, denoted below by $V_{L(R)}$  for tunneling through  the left (right) link.  When these are time dependent, the reflection, say to the left direction\aa{,} is then modified by factors of the form
\begin{align}
\int ^{t}dt' [V^{\dagger}_{L}(t) e^{i(\epsilon -\omega+i\Gamma)(t-t')}V^{}_{L}(t')+{\rm c.c.}]
\ ,
\label{int}
\end{align}
where $\Gamma$ is the width of the resonance formed on the dot (using $\hbar =1$). This form pertains to tunneling from the left lead to the dot, accomplished at time $t'$, followed by a time evolution of the electronic wave-function on the dot during the time interval $t-t'$, and then tunneling back to the left lead at time $t$.
One observes
that in the static case,  where $V^{\dagger}_{L}V^{}_{L}=1$,  the integral  (\ref{int}) yields the usual Breit-Wigner density of states on the dot, $2\Gamma/[(\omega-\epsilon)^{2}+\Gamma^{2}]$. For a
Rashba interaction that varies periodically with frequency $\Omega$,
 the reflection comprises multiple inelastic channels with emission and absorption of $n\Omega$ energy quanta, which shift the resonance above and below $\epsilon$. This complex modification of the reflection may differ for the opposite directions of the junction, leading to a net DC current.
Below we show that such a difference can indeed result from the Rashba interaction when the lengths of the two weak links %in Fig. \ref{sys}
are not identical.

%\vspace{0.3cm}

\noindent{\bf {\it 2. Details of the model.}}
\label{details}
The Rashba interaction in the  links is induced by  external electric fields, which can be polarized in various ways.  Here we focus on the simplest one of a longitudinal field (along the $\hat{\bf z}$ direction), whose amplitude oscillates with frequency $\Omega$ (see Fig. \ref{sys}).
The  Aharonov-Casher phase factor multiplying  the tunneling amplitude through a  link of  length $d$  along $\hat{\bf d}$ is then \cite{Aharony,moi}
\begin{align}
\exp[i\varphi^{}_{AC}(t)]=\exp[ik^{}_{\rm so}d\cos (\Omega t)\hat{\bf z}\times\hat{\bf d}\cdot\sig]\ ,
\end{align}
where  $\sig=(\sigma^{}_{x},\sigma^{}_{y},\sigma^{}_{z})$ is the vector of the Pauli matrices,  and $k_{\rm so}^{}$
is the  Rashba coupling. % which includes the strength of the generating electric field.
For the  geometry of  Fig.~\ref{sys}
the Aharonov-Casher phase factors are
\begin{align}
V^{}_{L(R)}(t)&=\cos[k^{}_{\rm so}d^{}_{L(R)}\cos(\Omega t)]\nonumber\\
&+i
\sin[k^{}_{\rm so}d^{}_{L(R)}\cos(\Omega t)]\sig\cdot\hat{\bf e}^{}_{L(R)}\ ,
\label{linp}
\end{align}
where $d_{L(R)}$ is the length of the link connecting the dot to the left (right) reservoir. For links along the $\hat{\bf x}-$direction (Fig.~\ref{sys}), the effective   magnetic fields
 created  by the Rashba interaction are along  $\hat{\bf e}^{}_{L}=-\hat{\bf y}$ and $\hat{\bf e}^{}_{R}=\hat{\bf y}$.
 %\begin{align}
%\hat{\bf e}^{}_{L(R)}=\hat{d}^{}_{L(R)x}\hat{\bf y}-\hat{d}^{}_{y}\hat{\bf x}\ ,
%\end{align}
%where $\hat{\bf d}^{}_{L(R)}$ is a unit vector that marks the link's direction.
% (Note that the $y$ components of ${\bf d}_{L}^{}$ and ${\bf d}^{}_{R}$ are identical.)

The entire junction
 is described by the Hamiltonian
\begin{align}
{\cal H}(t)={\cal H}^{}_{0}+{\cal H}^{}_{\rm tun}(t)\ ,
\end{align}
where the time-independent ${\cal H}^{}_{0}$ pertains to the decoupled system, i.e., two separate reservoirs and a quantum dot,
\begin{align}
{\cal H}^{}_{0}=
\sum_{\sigma}\epsilon d^{\dagger}_{\sigma}d^{}_{\sigma}+\sum_{{\bf k} ,\sigma}\epsilon^{}_{ k}c^{\dagger}_{{\bf k}\sigma}c^{}_{{\bf k}\sigma}+\sum_{{\bf p} ,\sigma}\epsilon^{}_{ p}c^{\dagger}_{{\bf p}\sigma}c^{}_{{\bf p}\sigma}\ .
\label{H0}
\end{align}
The first term in Eq.~(\ref{H0}) describes the decoupled dot, with
$d^{\dagger}_{\sigma}$ ($d^{}_{\sigma}$) being the creation (annihilation) operator of an electron of energy $\epsilon$ in the spin state $|\sigma\rangle$.
 The other two terms describe the decoupled electronic reservoirs,  assumed to comprise non-polarized free electrons. There, $
 c^{\dagger}_{{\bf k}\sigma} $ ($c^{}_{{\bf k}\sigma}$) creates (annihilates) a particle with energy $\epsilon_{k}$ ($\epsilon_p$),  momentum ${\bf k}$ (${\bf p}$),  and spin $\sigma$
 in the left (right) lead. The tunneling Hamiltonian reads 
\begin{align}
\label{Htun}
&{\cal H}^{}_{\rm tun}(t)=
\sum_{\alpha=L,R}{\cal H}^{\alpha}_{\rm tun}(t)%=\sum_\alpha{\cal H}^{\alpha=L(R)}_{\rm tun}(t)
\\
&=\sum_\alpha J^{}_{L(R)}\sum_{\sigma,\sigma'}\{[V^{}_{L(R)}(t)]^{}_{\sigma\sigma'}
%\nonumber\\&\times
\sum_{{\bf k}({\bf p})}c^{\dagger}_{{\bf k}({\bf p})\sigma}d^{}_{\sigma'}+{\rm H. c. }\}\ .
\nonumber
\end{align} 
The tunneling amplitudes,  characterized by the energy $J_{L(R)}$, are assumed to be
given by their values at the Fermi energy.

\noindent{\it{3. Currents in the time domain.}}
Within the Keldysh technique  \cite{Langreth,Jauho}   the particle current,  say into the left lead,  is  conveniently expressed in terms of the Green's function on the dot  \cite{Meir} $G_{dd}(t,t')$ (a matrix in spin space),
\begin{align}
I^{}_{L}(t)\equiv
\frac{d}{dt} \sum_{{\bf k}}\sum_{\sigma}&\langle c^{\dagger}_{{\bf k}\sigma}c^{}_{{\bf k}\sigma}\rangle
 = \int dt^{}_{1}{\rm Tr} \{\Sigma^{}_{L}(t,t^{}_{1})
G^{}_{dd}(t^{}_{1},t)\nonumber\\
&-G^{}_{dd}(t,t^{}_{1})\Sigma^{}_{L}(t^{}_{1},t)
 \}^{<}_{}\ ,
 \label{ILA}
\end{align}
where the angular brackets  denote quantum averaging. The superscript $<$ indicates the lesser Green's function, and $\Sigma^{}_{L}(t,t')$
is the self energy  due to the coupling of the dot to the left  reservoir,
\begin{align}
\Sigma^{}_{L}(t,t')=&J^{2}_{L}V^{\dagger}_{L}(t)g^{}_{L}(t,t')V^{}_{L}(t')\ ,
\label{ASEL}
\end{align}
where $g^{}_{L}(t,t')$ is the Green's function of the decoupled left  reservoir. Green's functions without a superscript represent  all three Keldysh Green's functions, the lesser,   and the retarded and advanced ones (marked by the superscripts $r$ and $a$).
The expression in Eq. (\ref{ILA}) can be worked out explicitly in the wide-band limit   \cite{Odashima,supplemental},  where  the densities of states in each of the reservoirs are approximated  by their value on the Fermi surface.
 The self energy $\Sigma_{L}^{r(a)}(t,t')$ is then
proportional to the unit matrix in spin space, with
\begin{align}
\Sigma^{r(a)}_{L}(t,t')&=\mp i\Gamma^{}_{L}\delta (t-t')\ ,
\label{rasel}
\end{align}
where
\begin{align}
\Gamma^{}_{L}&=\pi{\cal N}^{}_{L}J^{2}_{L}\ ,
\label{AgamL}
\end{align}
is the (partial) width of the resonance formed on the dot due to the coupling with the left reservoir  and  ${\cal N}^{}_{L}$ denotes the density of states of the left lead on the Fermi surface.  The total width of the resonance on the dot is $\Gamma=\Gamma_{L}+\Gamma_{R}$. The lesser self energy is a matrix in spin space,
\begin{align}
\Sigma^{<}_{L}(t,t')&=2i\Gamma^{}_{L}\int\frac{d\omega}{2\pi}e^{-i\omega(t-t')}f(\omega)V^{\dagger}_{L}(t)V^{}_{L}(t')\ .
\label{AlesSEL}
\end{align}
Here $f(\omega)=\{\exp[\beta(\omega-\mu)]+1\}^{-1}$ is the equilibrium Fermi distribution, with the inverse temperature $\beta$ and the chemical potential $\mu$ being identical for the two reservoirs.  [$\Sigma_{R}(t,t')$ is obtained from these expressions by changing $L$ to $R$.]

The explicit expression for $I_{L}(t)$ is found by applying the Langreth rules \cite{Langreth} to Eq. (\ref{ILA}),
\begin{align}
&I^{}_{L}(t)
=2\Gamma^{}_{L}{\rm Tr}\{-iG^{<}_{dd}(t,t) -\int\frac{d\omega}{2\pi}f(\omega)\nonumber\\
&\times
\int^{t}_{}dt^{}_{1}
[e^{-i(\epsilon-\omega+i\Gamma)(t^{}_{1}-t)}V^{\dagger}_{L}(t)V^{}_{L}(t^{}_{1})+{\rm c.c.}]\}\ .
\label{BIL}
\end{align}
The equal-time lesser Green's function  $-iG^{<}_{dd}(t,t)$, which yields the occupation of the dot,  is
\begin{align}
-iG^{<}_{dd}(t,t)&=\int\frac{d\omega}{\pi}f(\omega)[\Gamma^{}_{L}{\rm w}^{}_{L}(\omega,t)%\nonumber\\
+
\Gamma^{}_{R}{\rm w}^{}_{R}(\omega,t)]\ ,
\label{AG}
\end{align}
where
\begin{align}
&{\rm w}^{}_{L}(\omega,t)=\int^{t}_{}dt^{}_{1}\int ^{t}_{}dt^{}_{2}
e^{-i\omega (t^{}_{1}-t^{}_{2})}\nonumber\\
&\times
e^{-i(\epsilon-i\Gamma)(t-t^{}_{1})}e^{-i(\epsilon+i\Gamma)(t^{}_{2}-t)}
V^{\dagger}_{L}(t^{}_{1})V^{}_{L}(t^{}_{2})\ ,
\label{WA}
\end{align}
with an analogous expression for $w_{R}(\omega,t)$ (more details are given in Ref. \onlinecite{supplemental}). Thus, integrals of the form (\ref{int}) determine the explicit expressions for the current.

Using the  expansion \cite{AS}
\begin{align}
e^{i\zeta\cos(\phi)}=\sum_{n=-\infty}^{\infty}i^{n}J^{}_{n}(\zeta)e^{in\phi}\ ,
\end{align}
where $J_{n}(\zeta)$ in the Bessel function of integer order $n$,  one finds
(see Ref. \onlinecite{supplemental} for details)
\begin{align}
{\rm w}^{}_{L}(\omega,t)&=\Big
|J^{}_{0}(k^{}_{\rm so}d^{}_{L})D(\omega)\nonumber\\
&+\sum_{n=1}^{\infty}(-1)^{n}_{}J^{}_{2n}(k^{}_{\rm so}d^{}_{L})F^{}_{2n}(\omega,t)\Big |^{2}_{}\nonumber\\
&+
\Big |\sum_{n=0}^{\infty}(-1)^{n}_{}J^{}_{2n+1}(k^{}_{\rm so}d^{}_{L})F^{}_{2n+1}(\omega,t)\Big |^{2}_{}\ ,
\label{ww}
\end{align}
where
\begin{align}
D(\omega)=i/[\omega-\epsilon+i\Gamma]
\end{align}
is the Breit-Wigner resonance on the dot, and
\begin{align}
F^{}_{n}(\omega,t)=D(\omega-n\Omega)e^{in\Omega t}+D(\omega+n\Omega)e^{-in\Omega t}
\end{align}
is an even function of $\Omega$ that contains the contributions of the inelastic processes.
The second term on the right-hand side  of  Eq. (\ref{BIL}) is found in a similar fashion \cite{supplemental}. The particle current is then
\begin{align}
I^{}_{L}(t)=&4\Gamma^{}_{L}\Gamma^{}_{R}\int \frac{d\omega}{\pi}f(\omega)[{\rm w}^{}_{R}(\omega,t)-{\rm w}^{}_{L}(\omega,t)]\nonumber\\
&-2\Gamma^{}_{L}\int\frac{d\omega}{\pi}f(\omega)\frac{d{\rm w}^{}_{L}(\omega,t)}{dt}\ .
\label{ILC}
\end{align}
One notes that $I^{}_{L}(t)+I^{}_{R}(t)$ [the latter is obtained by interchanging $L$ with $R$ in Eq. (\ref{ILC})] equals the time derivative of $-{\rm Tr}\{iG^{<}_{dd}(t,t)\}$  [Eq. (\ref{AG})]  which is the occupation on the dot; i.e., charge is conserved in the junction. Note also that for $\Omega=0$ Eq. (\ref{ww}) becomes w$^{}_{L}(\omega)=[\cos^{2}(k^{}_{\rm so}d^{}_{L})+\sin^{2}(k^{}_{\rm so}d^{}_{L})]|D(\omega)|^{2}_{}=|D(\omega)|^{2}_{}$ which depends  neither on the length $d_{L}$ nor  on the spin-orbit coupling $k_{\rm so}$. In that case the first term on the right-hand side of Eq. (\ref{ILC}) vanishes, and there is no DC particle flow in an unbiased junction.

\begin{figure}[htp]
\includegraphics[width=5.5cm]{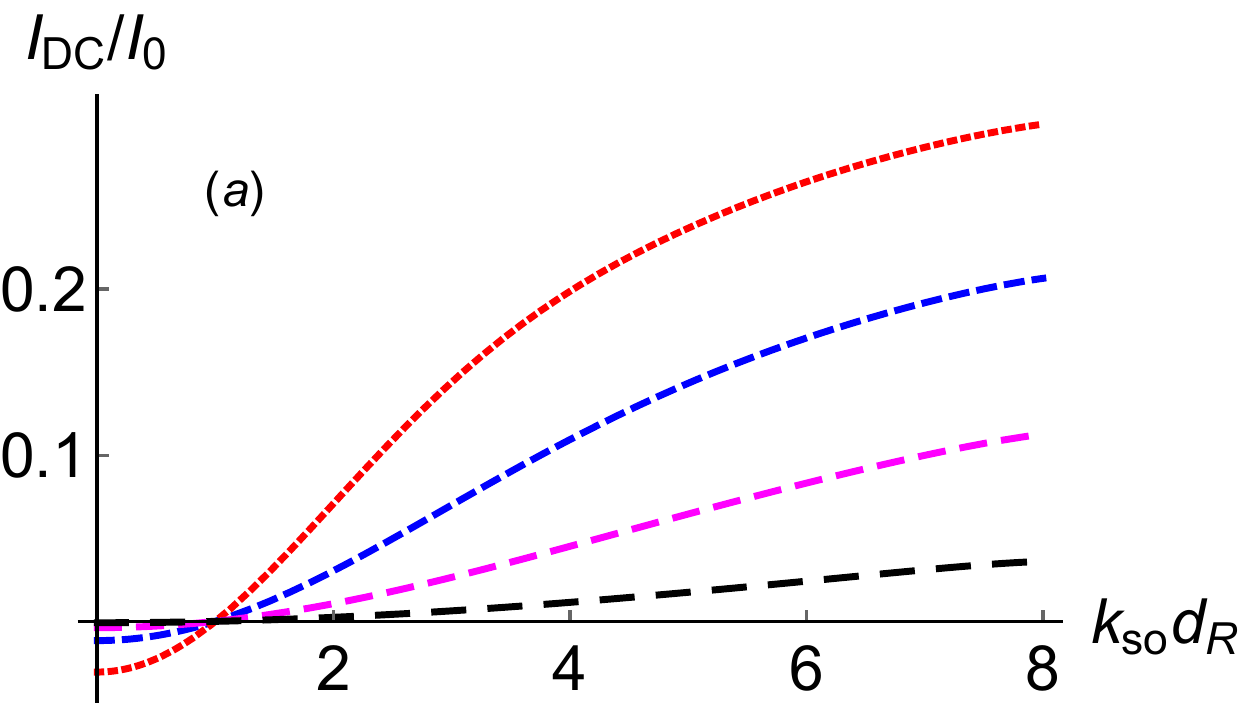}\\
\includegraphics[width=6cm]{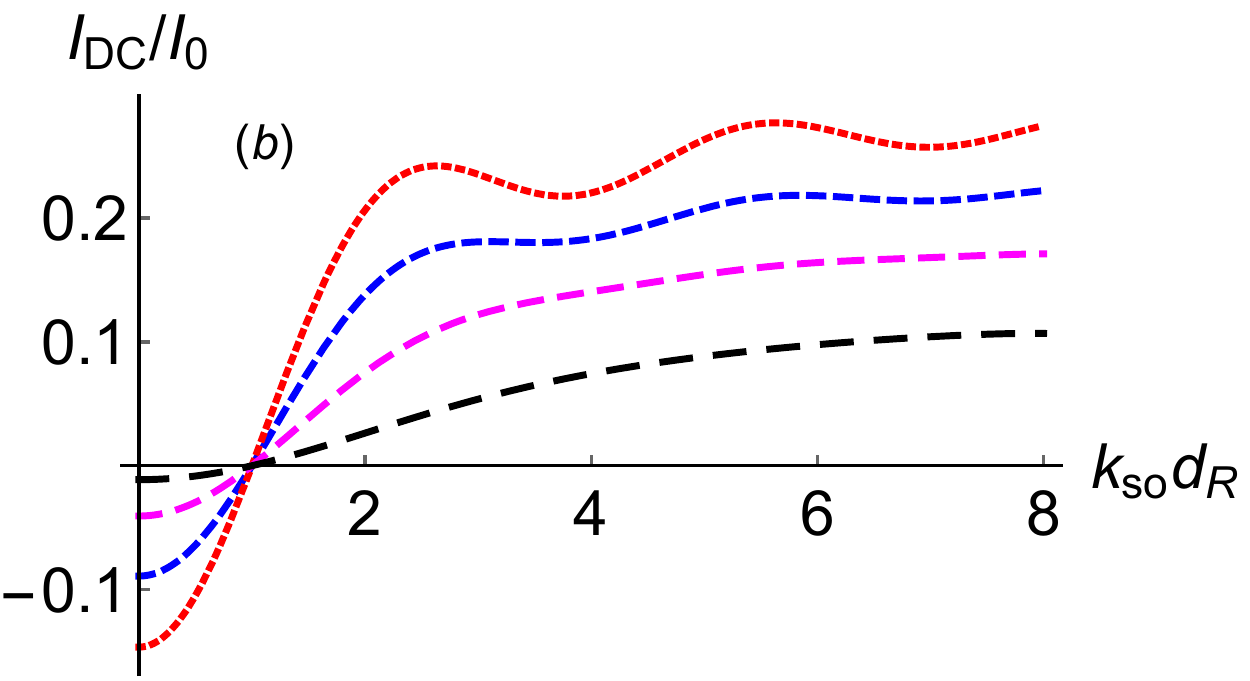}
\caption{The particle current $I_{\rm DC}$, normalized to $I^{}_{0}=(4\Gamma^{}_{L}\Gamma^{}_{R})/(\Gamma\pi\hbar)$,   calculated from Eq. (\ref{Idc})
as a function of $k^{}_{\rm so}d^{}_{R}$ for $k^{}_{\rm so}d^{}_{L}=1.0$,  $\epsilon-\mu=0.5\, \Gamma$;   (a) $\Omega =0.5\,\Gamma$, 
(b) $\Omega =2.0\,\Gamma$.  The increasing  dash lengths correspond to $\beta\Gamma$=10.0, 2.0, 1.0, and 0.5.}
\label{IDC}
\end{figure}

\noindent{\it{4. DC electromotive force generated by time-dependent Rashba interaction.}}
The current $I_{L}(t)$ comprises a static term, in addition to the time-dependent one. Obviously the derivative $d{\rm w}_{L}(t)/dt$ depends on time; but ${\rm w}_{L}$ (and similarly ${\rm w}_{R}$) contains a static term, ${\rm w}^{}_{L,{\rm s}}$, which takes a particularly simple form,
\begin{align}
{\rm w}^{}_{L,{\rm s}}(\omega)&=\sum_{n=-\infty}^{\infty}J^{2}_{n}(k^{}_{\rm so}d^{}_{L})|D(\omega-n\Omega)|^{2}
\ ,
\label{wLs}
\end{align}
which is even in $\Omega$.
As a result, the DC particle current through the junction is
\begin{align}
I^{}_{DC}&=\int \frac{d\omega}{\pi}4\Gamma^{}_{L}\Gamma^{}_{R}|D(\omega)|^{2}\nonumber\\
&\times\sum_{n=-\infty}^{\infty}f(\omega+n\Omega)[J^{2}_{n}(k^{}_{\rm so}d^{}_{R})-J^{2}_{n}(k^{}_{\rm so}d^{}_{L})]\ .
\label{Idc}
\end{align}
(The time-dependent parts of the currents can be found in Ref. \onlinecite{supplemental}.)
Figure \ref{IDC} portrays the current   {\it vs.} $k^{}_{\rm so}d^{}_R$ at a fixed value of $k^{}_{\rm so}d^{}_L$, as calculated from Eq. (\ref{Idc}) for several values of  temperature and $\Omega$.  The  oscillations, at low temperatures, reflect the oscillatory length-dependence of the effect of the Rashba interaction \cite{Aharony}. 
These oscillations  disappear gradually as the temperature is raised. At low temperatures and large $\Omega$'s, $I_{DC}$ is dominated by the oscillations of the zeroth and first order Bessel functions.

The appearance of  DC  electronic charge transport  in the absence of a bias voltage across  the device is a manifestation of an electromotive force acting in the electric circuit. In our case, the force relies on the electronic spin, and drives the electron\st{s'} flow through a junction subject to a time-dependence spin-orbit interaction. The driving occurs due to the fermionic  nature of the electrons which imposes  constraints on the inelastic spin-scattering induced by the time-dependent Rashba interaction:
some of the inelastic scattering channels  become
partly blocked due to the Pauli principle. This is why the unitarity of spin transmission, which would hold if all inelastic transmission channels would be equally open, is broken \cite{sumrule}.
%Indeed,   examining Eq. (\ref{Idc}) one observes that \aa{when $\Omega=0$,  $f(\omega+n\Omega)$ is independent of $n$, and then
%  $I_{DC}$ would vanish (since $\sum_{n}J^{2}_{n}(x)=1$ \cite{AS})}.
The peculiar photovoltaic effect discussed above  manifests itself in inhomogeneous devices with a well-defined direction of the inhomogeneity along the direction of the  current flow. The single-dot tunneling device studied here, in which the reflection asymmetry is generated by the different lengths of the links (in conjunction with the Rashba coupling), is an example of such an inhomogeneity.

%%%%%%%%%%%%%%%%%%%%%%%%%%%%%%%%%%%%%%%%%%%%%%%%%%%%%%%%%%%%
%%%%%%%%%%%%%%%%%%%%%%%%%%%%%%%%%%%%%%%%%%%%%%%%%%%%%%%%%%%%

One may propose a simple procedure to measure the spin-orbit-induced electromotive force. Suppose that the spin-orbit coupled weak link (which contains the dot) is an element of an {\it open} electric circuit. Then the spin-driven electromotive force would lead to an accumulation of extra  charges of opposite signs on the two terminals, and in turn to the building up of a voltage drop across the junction. A steady-state configuration would then emerge, provided that the ohmic current generated by this voltage drop  compensates the DC current due to the electromotive force induced by the Rashba interaction.

Thus, a novel photovoltaic effect can be predicted: a microwave field applied to a gate electrode (see Fig. \ref{sys}) induces a voltage drop across the junction.
A simple analytical estimate of the
 voltage signal can easily be obtained for a weak Rashba coupling, 
$k^{}_{\rm so}d^{}_{L(R)}\ll 1$: In the absence of the spin-orbit interaction, a bias voltage $V$ on the junction gives rise to a DC particle current  \cite{Jauho}
\begin{align}
I^{}_{\rm DC}=\frac{4\Gamma^{}_{L}\Gamma^{}_{R}}{\pi\hbar}\frac{eV}{\epsilon^{2}_{0}+\Gamma^{2}}\sim
\frac{4\Gamma^{}_{L}\Gamma^{}_{R}}{\pi\epsilon^{2}_{0}}\frac{eV}{\hbar}\ ,
\label{IDC0}
\end{align}
where $\epsilon_{0}^{}\equiv \epsilon-\mu $ is assumed to be much larger than $\Gamma $ in the last step ($\hbar$ was re-introduced into the expressions for the following estimates).
On the other hand, according to Eq. (\ref{Idc}) at small spin-orbit coupling, the oscillating Rashba interaction generates the zero-temperature particle  DC current 
\begin{align}
&I^{}_{\rm DC}\Big |^{}_{T=0}=\frac{4\Gamma^{}_{L}\Gamma^{}_{R}}{\pi\hbar\Gamma}\frac{k^{2}_{\rm so}[d^{2}_{L}-d^{2}_{R}]}{2}\\
&\times\Big [\frac{1}{2}\Big (
\arctan\frac{\epsilon^{}_{0}+\Omega}{\Gamma}
+
\arctan\frac{\epsilon^ {}_{0}-\Omega}{\Gamma}\Big )-\arctan\frac{\epsilon^ {}_{0}}{\Gamma}\Big ]
\ ,
\nonumber
\end{align} 
which becomes, for $\epsilon^{}_{0}\gg\Gamma$ and $\Omega<\epsilon^{}_{0}$
\begin{align}
&I^{}_{\rm DC}\Big |^{}_{T=0}\sim
\frac{4\Gamma^{}_{L}\Gamma^{}_{R}}{\pi\epsilon^{2}_{0}}k^{2}_{\rm so}[d^{2}_{R}-d^{2}_{L}]\frac{(\hbar\Omega)^{2}}{2\hbar\epsilon^{}_{0}}\ .
\label{IT0}
\end{align}
Thus, the voltage drop $V^{}_{\rm em}$ generated by the electromotive force is
\begin{align}
V^{}_{\rm em}=k^{2}_{\rm so}(d^{2}_{R}-d^{2}_{L})(\hbar\Omega)^{2}/(2e\epsilon^{}_{0})\ .
\end{align}
Similar considerations for the oscillatory length-dependence pertain to the case where the  Rashba interaction is induced by mechanical vibrations of the nanowire forming the link \cite{RS2013, MJ2018}. 

Our model system could be implemented by, e.g., three in-line, side-gated InAs nanowires \cite{Scherubl}. The left and right nanowires of length $d_{L}$ and $d_{R}$ would serve as weak links and be in tunneling contact with, respectively, the source- and drain electrodes as well as with the short, central nanowire, which would serve as a quantum dot. One of the two gates on either side of the weak links would be excited by a microwave field that creates an AC gate voltage, $V_{ac}\cos(\Omega t)$ \cite{Nowack},  while a static voltage on the the two gates on either side of the quantum dot would be used to tune the energy levels in the dot.

With a distance of $\sim$200 nm between the side gates \cite{Scherubl} a microwave-generated amplitude of  $V_{ac} = 1$ V on the side gates would produce a transverse electric field amplitude of $\sim$50 kV/cm in the wires, corresponding to a Rashba parameter $\alpha_R = \hbar^2 k_{\rm so}/m^* \sim 50\, {\rm meV} \cdot$\AA\,  \cite{Zunger} and, using $m^* = 0.023 m_e$, a Rashba coupling $k_{\rm so} \sim 2\cdot 10^{-3}\,$(nm)$^{-1}$ in the weak links. Assuming $d_{L} \sim d_{R} \sim 250\,$nm and microwave frequency of $2\pi \times 100\,$ GHz (so that $\hbar \Omega$ is of the order of the energy level $\epsilon_0\sim 1\,$meV [with respect to the chemical potential]) one finds $V_{\rm em}\sim 5\ \mu$V; thus the photovoltaic voltage in response to the microwave field seems  to be measurable. Using the same parameter estimates the particle current, Eq.  (\ref{IT0}),  is $\sim 5\cdot 10^{7}$ s$^{-1}$, corresponding to a charge current $\sim 10$ pA.

\noindent{\it{5. Summary.}}
We have found that the spin-orbit (Rashba) interaction confined to an electric weak link, which -- when static -- has no significant effect on DC transport of two-terminal devices,  may act as a source of DC currents
 when generated by a periodic electric field. This electric field renders the Rashba interaction time dependent, breaking the unitarity of the spin transmission by generating inelastic transmission channels. We have shown that this loss of unitarity appears as additional contributions to the backscattering [see Eqs. (\ref{int}) and (\ref{WA})]. An estimate of the generated voltage drop in an open circuit suggests that it can be detected experimentally.

The effect we find is due to modifications of the probabilities for electron reflections, which are different for electrons approaching the junction from opposite directions; nonetheless, it is not related to quantum pumping \cite{Thouless}. The origin of the latter are different time-dependent {\em phases} of the instantaneous reflection amplitudes \cite{Avron}, whereas a straightforward calculation of the instantaneous scattering matrix
for the junction illustrated in Fig. \ref{sys}
shows that the reflection amplitudes do not depend on time. This is because $V^{\dagger}_{L}(t)V^{}_{L}(t)=1$ due to the unitarity of the Aharonov-Casher phase factor.  In our case, the reflections are modified by Aharonov-Casher phase factors at different times, and necessitate   the inclusion of the inelastic dynamics on the dot. % [see Eq. (\ref{int})].

%%%%%%%%%%%%%%%%%%%%%%%%%%%%%%%%%%%%%%%%%%%%%%%%%%%%%%

\begin{acknowledgments}
This research was  partially supported by the Israel Science Foundation (ISF), by the infrastructure program of Israel Ministry of Science and Technology under contract 3-11173, and  by the Pazy Foundation. We acknowledge the hospitality of the PCS at IBS, Daejeon, Korea, and Zhejiang University, Hangzhou, China,  where part of this work was  supported by
IBS funding number
(IBS-R024-D1).
\end{acknowledgments}

%%%%%%%%%%%%%%%%%%%%%%%%%%%%%%%%%%%%%%%%%%%%%%%%%%%%%%
%%%%%%%%%%%%%%%%%%%%%%%%%%%%%%%%%%%%%%%%%%%%%%%%%%%%%%

%%%%%%%%%%%%%%%%%%%%%%%%%%%%%%%%%%%%%%%%%%%%%%%%%%%%%%

%%%%%%%%%%%%%%%%%%%%%%%%%%%%%%%%%%%%%%%%%%%%%%%%%%%%%%


\begin{thebibliography}{99}
\bibitem{sumrule}
Trivially, all channels are equally open when $\Omega=0$,  in which case $f(\omega+n\Omega)$ is independent of $n$ and $I_{DC}$, given by Eq. (\ref{Idc}), vanishes (since $\sum_{n}J^{2}_{n}(x)=1$ \cite{AS}).
\end{thebibliography}
\end{document}